\documentclass[a4paper,envcountsame,envcountsect]{llncs}

\usepackage{amsfonts}

\begin{document}

\title{On the Solution of Graph Isomorphism by Dynamical Algorithms}

\author{
Marats Golovkins
\thanks{Supported by the Latvian Council of Science, grant No. 01.0354. The paper 
was prepared while visiting LIAFA, Universit\'e Paris 7.}
}

\institute{
 Institute of Mathematics and Computer Science,
 University of Latvia\\ Rai\c na bulv. 29, Riga, Latvia\\
\email{marats@latnet.lv}
}

\maketitle


\begin{abstract}
In the recent years, several polynomial algorithms of a dynamical nature have been proposed
to address the graph isomorphism problem (\cite{GN 02}, \cite{SJC 03}, etc.).
In this paper we propose a generalization of an approach exposed in \cite{GN 02} and
find that this dynamical algorithm is covered by a combinatorial approach. It is possible to 
infer that polynomial dynamical algorithms addressing graph isomorphism are covered by suitable 
polynomial combinatorial approaches and thus are tackled by the same weaknesses as the last ones.

\end{abstract} 

\section{Introduction}
\label{intro}
In their paper, Gudkov and Nussinov proposed a new approach to analyze graphs. Suppose a graph has $n$ vertices. Then each vertex of the graph
is viewed as a point-mass in an $n-1$ - dimensional Euclidean space. The point-set is arranged into a symmetrical initial configuration and afterward
interacted by mutual attraction and repulsion forces, with attraction force acting on every pair of point-masses corresponding to vertices having an edge 
between them. So the point-set is subjected to distortion. It is possible to simulate the dynamics numerically,
thus computing the future coordinates of the point-masses.
Given two graphs, each of them is embedded into separate $n-1$ - dimensional space. If after some time the two point-sets are no longer congruent,
the two graphs are not isomorphic. However, testing the congruence of two point-sets in a $d$-dimensional space is an open problem as well, being at least as hard
as graph isomorphism itself \cite{A 98}. Eventually the authors have chosen the following approach. A set of $n(n-1)/2$ numbers for each graph 
is assigned being a set of mutual distances between the points in the relevant point-set at a certain moment of time.
If the distance sets for the two graphs do not match, the two graphs are not isomorphic. The authors also conjectured that if the distances do match, the graphs
are isomorphic. However, as pointed out in \cite{SJC 03}, the distances are the same for two non-isomorphic strongly regular graphs with the same parameters.

Essentially Gudkov and Nussinov apply their approach to give an answer to the {\em graph coding problem}, i.e., find a polynomial time algorithm which
assigns an integer number to each graph, so that two graphs share the same number if and only if they are isomorphic. However, graph
coding is not necessary to solve graph isomorphism, as graph {\em vertex classification} up to automorphism partition classes is sufficient \cite{RC 77}.
Informally, the aim of vertex classification is to assign a local code to each vertex in the graph. The codes for two 
vertices in the same graph would be the same if and only if the two vertices play an equivalent role in the graph.

In this paper, we shall use the idea of dynamical evolution to give a graph vertex classification algorithm, hereinafter referred as algorithm A1.
Informally, given a graph on $n$ vertices, it
is embedded into $n$-dimensional space. Now the initial symmetric configuration is defined by an $n\times n$ identity matrix, which rows form the coordinates
of the points. Afterward the point-set is subjected to a dynamical evolution determined by an adequate definition of attraction and repulsion 
forces acting on point-masses. If at some moment of time the coordinates of two points in the system are not permutation equivalent, the corresponding
vertices are not in the same class of automorphism partition.

In \cite{SJC 03} the authors show that the original \cite{GN 02} algorithm does not work on strongly regular graphs and explain that 
by the combinatorial properties of strongly regular graphs. On the other hand, the algorithm A1 works on many pairs of strongly regular graphs, 
including those exposed in \cite{SJC 03}. However, it is still possible to indicate instances of graphs where the A1 algorithm fails.

In Section \ref{prel}, we introduce the definitions and prove the basic theorems used in this paper.
Section \ref{dynamic} presents the dynamical system which forms the basis for the partitioning algorithm A1.
Section \ref{algo} presents the algorithm A1, establishes its complexity, shows that is is covered by a polynomial combinatorial algorithm and indicates
a counterexample, where the algorithm fails to distinguish between two non-isomorphic graphs.

The fact that polynomial dynamical algorithms (at least in the presented case) are covered by polynomial combinatorial algorithms
is the main result of this paper.

In the final Section \ref{conj} it is shown
that A1 may be extended to a yes-no-don't-know type algorithm A2, which either gives a correct answer or says nothing.

\section{Preliminaries}
\label{prel}

\begin{definition}
An (undirected) {\em graph} is an ordered pair of disjoint sets $(V,\Gamma)$, where $\Gamma\subseteq V\times V$ is an anti-reflexive and symmetric relation.
Elements of the set $V$ are called {\em vertices} of the graph.
If $v_1,v_2\in V$, $v_1\Gamma v_2$ and $v_2\Gamma v_1$, then a set $\{v_1,v_2\}$ is called an {\em edge} of the graph.
\end{definition}

Thus every graph can be denoted as an ordered pair of disjoint sets $G=(V,E)$, where $V$ is the set of vertices and $E$ - the set of edges of the graph.
We denote by $|G|$ the number of vertices in the graph $G$.

Graphs $G_1=(V_1,E_1)$ and $G_2=(V_2,E_2)$ are called {\em separate}, if $V_1\cap V_2=\emptyset$.

\begin{definition}
An {\em adjacency matrix} of a graph $G=(V,E)$ on $n$ vertices is an $n\times n$ 0,1-matrix $H$, such that $h_{i,j}=1$ iff $\{v_i,v_j\}\in E$.
\end{definition}

\begin{definition}
A path on a graph $G=(V,E)$ is a sequence $v_1,v_2,\dots,v_m$, $v_i\in V$, such that $\{v_1,v_2\}$, $\{v_2,v_3\},\dots,\{v_{m-1},v_m\}$ are edges of the graph.
\end{definition}

\begin{definition}
Two vertices $v_1,v_2$ are called $l$-connected in a graph $G$, if exists a path $v_1,\dots,v_2$ of $l+1$ elements on the graph $G$.
Two vertices are called connected, if exists $l\geq0$ such that the vertices are $l$-connected.
Otherwise the two vertices are called disconnected.
By default, every vertex is considered to be connected to itself.

A graph $G=(V,E)$ is called {\em connected}, if all vertices in the graph are connected.
Otherwise the graph is called disconnected.

A graph vertex $v$ is of degree $k$, $\deg(v)=k$, iff there are $k$ different edges that contain $v$.

A {\em degree partition} $D(G)$ of a graph $G=(V,E)$ is formed by an equivalence relation among the graph vertices $\{(v_1,v_2)\in V^2\ |\ \deg(v_1)=\deg(v_2)\}$.
\end{definition}

%
\begin{definition} 
A {\em complement} of a graph $G=(V,E)$ is a graph $\overline{G}=(V,\overline{E})$, where
$\overline{E}=\big\{\{v_1,v_2\}\ \big|\ \{v_1,v_2\}\notin E\ \&\ v_1\neq v_2\ \&\ v_1,v_2\in V\big\}$.
\end{definition}

\begin{theorem}
\label{disc_con}
If a graph is disconnected, then its complement is connected.
\end{theorem}
\begin{proof}
Let $G=(V,E)$ be a disconnected graph and $v_1,v_2$ - two vertices of the graph.
If $v_1,v_2$ are disconnected in the graph $G$, they form an edge $\{v_1,v_2\}$ in a graph $\overline{G}$ and thus are connected in $\overline{G}$.

Suppose $v_1,v_2$ are connected in $G$. Since $G$ is disconnected, exists a vertex $v_0$, which is not connected to $v_1$ in $G$. Hence 
$v_0$ is not connected to $v_2$ in $G$ as well. Therefore the vertices $v_0,v_1,v_2$ form edges $\{v_1,v_0\}$ and $\{v_0,v_2\}$ in $\overline{G}$.
Thus $v_1,v_0,v_2$ is a path on the graph $\overline{G}$ and the vertices $v_1,v_2$ are connected in $\overline{G}$.

Hence the graph $\overline{G}$ is connected.\qed
\end{proof}

\begin{definition}
We shall call a graph {\em doubly connected \cite{L 00}} if both the graph and its complement are connected.
\end{definition}

We can apply a bijection $V_1\stackrel{\beta}{\longrightarrow}V_2$ on the set of edges of a graph $G=(V_1,E_1)$, i.e.,
$\{v_1,v_2\}\beta\stackrel{\scriptscriptstyle {\rm def}}{=}\{v_1\beta,v_2\beta\}$, hence
$E_1\beta=\big\{\{v_1\beta,v_2\beta\}\ \big|\ \{v_1,v_2\}\in E_1\big\}$.

\begin{definition}
Two graphs $G_1=(V_1,E_1)$, $G_2=(V_2,E_2)$ are called {\em isomorphic} if exists a bijection $V_1\stackrel{\beta}{\longrightarrow}V_2$
such that $E_1\pi=E_2$. The bijection $\beta$ is then called {\em an isomorphism}.
\end{definition}

Thus {\em graph isomorphism problem} for two graphs $G_1$ and $G_2$ means finding whether exists an isomorphism between these two graphs.

A permutation $\pi$ on $V$ is a bijection $V\stackrel{\pi}{\longrightarrow}V$.

\begin{definition}
A permutation $\pi$ on $V$ is called {\em an automorphism} of a graph $G=(V,E)$, if $E\pi=E$.
\end{definition}

We denote by $A(G)$ the set of all automorphisms of a graph G.
The following lemma is straightforward.

\begin{lemma}
\label{autlem}
Given a graph $G=(V,E)$ and an automorphism $\pi\in A(G)$,
for all $v_1,v_2\in V$ $v_1,v_2$ are $1$-connected iff $v_1\pi,v_2\pi$ are $1$-connected.
\end{lemma}

\begin{corollary}
\label{degaut}
Given a graph $G=(V,E)$ and an automorphism $\pi\in A(G)$, for all $v\in V$ $\deg(v)=\deg(v\pi)$.
\end{corollary}

The automorphism relation $\alpha$ among the vertices of the graph is defined as follows.

\begin{definition}
For two vertices $v_1,v_2\in V$ of a graph $G=(V,E)$, $v_1\alpha v_2$ iff exists an automorphism $\pi\in A(G)$ such that
$v_1\pi=v_2$.
\end{definition}

Since $\alpha$ is an equivalence relation, it spans the set of vertices $V$ into a set of equivalence classes, called {\em automorphism partition} of
the vertices of a graph.

Thus {\em automorphism partitioning problem} for a given graph is to find a partition of the graph's vertices into automorphism equivalence classes.

We denote by $P(G)$ the automorphism partition of a graph $G$.
It is quite evident that given a graph $G$, $P(G)=P(\overline{G})$.
By Corollary \ref{degaut}, if two vertices are in the same class of automorphism partition, they are in the same class of degree partition.

\begin{definition}
Two partitions of $P(V_1),P(V_2)$ of the vertices of separate graphs $G_1=(V_1,E_1)$ and $G_2=(V_2,E_2)$ are called {\em equivalent}, if 
exists a bijection $V_1\stackrel{\beta}{\longrightarrow}V_2$ such that $P(V_1)\beta=P(V_2)$.
\end{definition}

Two isomorphic graphs have equivalent automorphism partitions, the opposite is not necessarily true.

In \cite{RC 77} it is shown that graph automorphism partitioning problem and graph isomorphism problem are polynomially equivalent.
As mentioned in Section \ref{intro}, algorithm A1 is a vertex classification algorithm toward automorphism partitioning classes.
For the reasons exposed in Section \ref{algo}, A1 is applied only on doubly connected graphs.
Therefore it is necessary to state the following theorem.

\begin{theorem}
\label{reduct}
The graph isomorphism problem is polynomially reducible to the automorphism partitioning problem of
a doubly connected graph.
\end{theorem}
\begin{proof}
Assume a polynomial automorphism partitioning algorithm is given.
Consider two separate graphs $G_1=(V_1,E_1)$ and $G_2=(V_2,E_2)$, $|G_1|=|G_2|=n>1$.
Assume that graphs have equivalent degree partitions, otherwise the graphs are not isomorphic.
Further assume that both $G_1$ and $G_2$ are connected, otherwise consider their complements
$\overline{G_1}$ and $\overline{G_2}$.
Select a vertex $\sigma_0\in V_1$ with a maximal degree in $G_1$, $\deg(\sigma_0)=d_0$, and for each vertex $\tau\in\{v\in V_2\ |\ \deg(v)=d_0\}$
form $G_\tau=(V,E_\tau)$, where $V=V_1\cup V_2$ and $E_\tau=E_1\cup E_2\cup\big\{\{\sigma_0,\tau\}\big\}$.
$G_\tau$ is connected.
The graph $\overline{G_\tau}$ is connected. (Proof is similar to that of Theorem \ref{disc_con}.)
Therefore $G_\tau$ is doubly connected.

Now for every $\tau$ perform the automorphism partitioning on $G_\tau$.
If exists $\tau$ such that $\sigma_0\alpha\tau$, then $G_1$ and $G_2$ are isomorphic, otherwise the graphs are not isomorphic. 
(Lemma \ref{i_at}.)

The number of performed automorphism partitionings is bounded by $n$, hence the reduction is polynomial. 
\qed
\end{proof}

The following lemma certifies the correctness of the reduction algorithm described in the proof of Theorem \ref{reduct}.
\begin{lemma}
\label{i_at}
The graphs $G_1=(V_1,E_1)$ and $G_2=(V_2,E_2)$ are isomorphic iff exists $\tau\in V_2$ such that $\sigma_0\alpha\tau$ holds for $G_\tau$.
\end{lemma}
\begin{proof}
Suppose that exists $\tau\in V_2$ such that $\sigma_0\alpha\tau$ holds for $G_\tau$.
Hence exists $\pi\in A(G_\tau)$ such that $\sigma_0\pi=\tau$.
Vertices $\sigma_0,\tau$ are the only vertices with degree $d_0+1$ in $G_\tau$. Therefore they are the only members of an automorphism class
in $P(G_\tau)$. Hence $\tau\pi=\sigma_0$.
Since $V_1$ is connected, it is possible to construct a sequence $\{\sigma_0\}=X_0\subset X_1\subset\dots\subset X_{n-1}=V_1$, 
where a set $X_{i+1}$ is formed by adding to $X_i$ a vertex in $V_1\setminus X_i$ which is $1$-connected to a vertex in $X_i$.
Let us prove that $\forall i\geq0\ X_i\pi\subset V_2$.
Proof is by induction. As for induction basis, $\sigma_0\pi=\tau\in E_2$.
Now, for induction step, assume that $X_i\pi\subset V_2$.
Take the vertex $v\in X_{i+1}\setminus X_i$. Consider a vertex $x\in X_i$, which is $1$-connected to $v$. Since
$x\pi\in V_2$, by Lemma \ref{autlem},
either $v\pi=\sigma_0$ or $v\pi\in V_2$. However $\neg(v\alpha\sigma_0)$, so $v\pi\in V_2$ and $X_{i+1}\pi\subset V_2$.
Now, $V_1\pi\subset V_2$. Since $\pi$ is a permutation and $|V_1|=|V_2|$, $V_1\pi=V_2$. Furthermore, that leads to $V_2\pi=V_1$.
Since $E_\tau\pi=E_\tau$, $E_1\pi\cup E_2\pi=E_1\cup E_2$. Since $E_1\cap E_2=E_1\pi\cap E_2\pi=E_1\cap E_1\pi=E_2\cap E_2\pi=\emptyset$,
$E_1\pi=E_2$ and $E_2\pi=E_1$. Hence $\pi$ is an isomorphism from $G_1$ to $G_2$.

Suppose $G_1$ is isomorphic to $G_2$. Then exists an isomorphism $V_1\gamma=V_2$, $E_1\gamma=E_2$. Define a permutation $\pi$ on $V_1\cup V_2$ as
$\pi=\gamma\cup\gamma^{-1}$. Now $V_1\pi=V_2$, $V_2\pi=V_1$, $E_1\pi=E_2$ and $E_2\pi=E_1$.
Take $\tau=\sigma_0\pi$. By construction, $\tau\pi=\sigma_0$. Consider the graph $G_\tau$. For the set of the graph's vertices, 
$V\pi=(V_1\cup V_2)\pi=V_1\pi\cup V_2\pi=V_1\cup V_2=V$.
For the set of the graph's edges, $E_\tau\pi=\Big(E_1\cup E_2\cup\big\{\{\sigma_0,\tau\}\big\}\Big)\pi=E_1\pi\cup E_2\pi\cup\big\{\{\sigma_0,\tau\}\pi\big\}
=E_1\cup E_2\cup\big\{\{\sigma_0,\tau\}\big\}=E_\tau$. Hence $\pi$ is an automorphism of $G_\tau$. Therefore $\sigma_0\alpha\tau$.
\qed
\end{proof}

\section{Dynamical System and Its Basic Properties}
\label{dynamic}
In this section we define the dynamical system and prove basic properties of the system upon which the partitioning algorithm A1 will depend on.

We shall take use of some definitions and theorems from the theory of real analytic functions.
For detailed exposition, see \cite{KP 02}.

Suppose a graph on $m$ vertices is given and its adjacency matrix is $H=(h_{ik})$. The basic idea behind is to consider the graph as a physical
system, embedded into $m$-dimensional space. The reason why dimension should coincide with the number of vertices will be explained in Section \ref{algo}.
Vertices of a graph are points that enjoy mutual attraction and repulsion forces. Every two points are interacted by repulsion force, whereas
every two points corresponding to 1-connected vertices are also interacted by attraction force. The mass of any point is defined to be $1$.

The coordinates of $m$ points form an $m\times m$ matrix $X$, where the $i$-th row of the matrix denotes the coordinates of the $i$-th point,
$x_i=(x_{i1},x_{i2},\dots,x_{im})$.

Repulsion force $Fr_{ik}$ between any two points $x_i$ and $x_k$ is chosen
to be inversely proportional to the distance $\|x_i-x_k\|=\sqrt{\sum\limits_{l=1}^m(x_{il}-x_{kl})^2}$ between the two points,
$\|Fr_{ik}\|=1/\|x_i-x_k\|$, whereas attraction force $Fa_{ik}$ between the points connected by an edge
is chosen to be directly proportional, $\|Fa_{ik}\|=\|x_i-x_k\|$.

In principle, it is possible to define the forces in a different way, say, $\|Fr_{ik}\|=1/\|x_i-x_k\|^2$
or $\|Fr_{ik}\|=1$. 
In this paper, we did not attempt to investigate the impact of choosing particular attraction or repulsion force, however it seems
that attraction and repulsion forces should be linearly independent.

So attraction force acting on a point $x_i$ due to its interaction with a point $x_k$ is
\begin{equation} 
Fa_{ik}=-h_{ik}(x_i-x_k),
\end{equation}
whereas repulsion force
\begin{equation}
{Fr}_{ik}=\frac{x_i-x_k}{\|x_i-x_k\|^2}.
\end{equation}
The total force acting on $x_i$ from $x_k$ is
\begin{equation}
F_{ik}=Fr_{ik}+Fa_{ik}=(x_i-x_k)\bigg(\frac{1}{\|x_i-x_k\|^2}-h_{ik}\bigg).
\end{equation}
Hence $F_{ik}=-F_{ki}$ and $F_{ii}=0$.
So the total force acting on $x_i$ is
\begin{equation}
F_i=\sum\limits_{k=1}^mF_{ik}=\sum\limits_{\scriptstyle k=1\atop\scriptstyle k\neq i}^m(x_i-x_k)\bigg(\frac{1}{\|x_i-x_k\|^2}-h_{ik}\bigg).
\end{equation}
In our setting, $$\frac{d^2x_{i}(t)}{dt^2}=F_i,$$
hence the dynamics is described by the following autonomous system of differential equations;
\begin{equation}
\label{difeq}
\scriptstyle{i,j}\displaystyle\Bigg\{_1^m x_{ij}^{\prime\prime}=
     \sum\limits_{\scriptstyle k=1\atop\scriptstyle k\neq i}^m(x_{ij}-x_{kj})\bigg(\frac{1}{\|x_i-x_k\|^2}-h_{ik}\bigg).
\end{equation}

Let $\mathcal{M}$ be a set of $m\times m$ real matrices, whereas $\mathcal{M^+}\subset\mathcal{M}$ a set of matrices where no two rows in the matrix are the same.
The right hand side of (\ref{difeq}) may be regarded as a function $F_H:\mathcal{M^+}\longrightarrow\mathcal{M}$. Hence the system (\ref{difeq}) may be rewritten
as 
\begin{equation}
\label{difeq1}
X^{\prime\prime}=F_H(X).
\end{equation}
By making a standard reduction to first-order differential equations, we obtain
\begin{equation}
\label{difeqfo}
\left\{
\begin{array}{l}
X^\prime=Y\\
Y^\prime=F_H(X).
\end{array}
\right.
\end{equation}
Given $B\in\mathcal{M^+},V\in\mathcal{M}$, initial conditions
for (\ref{difeqfo}) are specified as
\begin{equation}
\label{icond}
\left\{
\begin{array}{l}
X(t_0)=B\\
Y(t_0)=V.
\end{array}
\right.
\end{equation}
The system (\ref{difeqfo}) is in fact a Hamiltonian system, that is,
\begin{equation}
\begin{array}{l}
\scriptstyle \ \ \ \ \ m\vspace{-2 mm}\\
\scriptstyle{i,j}
\left\{
\begin{array}{l}
\ \ x_{ij}^{\prime}=\displaystyle\frac{\partial\mathcal{H}}{\partial y_{ij}}
\vspace{1 mm}
\\
\ \ y_{ij}^{\prime}=\displaystyle-\frac{\partial\mathcal{H}}{\partial x_{ij}},\\
\end{array}
\right.\vspace{-2 mm}\\
\scriptstyle \ \ \ \ \ 1
\end{array}
\end{equation}
where $y_{ij}$ are elements of $Y$ and
\begin{eqnarray}
\mathcal{H}&=&E_k(Y)+E_p(X)\ ,\\
E_k(Y)&=&\frac{1}{2}\sum\limits_{r,s=1}^m y_{rs}^2\ ,\\
E_p(X)&=&-\frac{1}{2}\sum\limits_{\scriptstyle r,s=1\atop\scriptstyle r<s}^m\left(
\log\|x_r-x_s\|^2-h_{rs}\|x_r-x_s\|^2
\right).
\end{eqnarray}
Functions $E_k(Y)$ and $E_p(X)$ are frequently referred as kinetic and potential energy, respectively.
The system (\ref{difeqfo}) being an autonomous Hamiltonian system, energy conservation law (\cite{HS 74}, p.292; \cite{TP 63}, p.475) is applicable, yielding
\begin{equation}
\label{energy}
E_k(Y)+E_p(X)=E_0,
\end {equation}
where $E_0$ is a constant value; using initial conditions (\ref{icond}) one can compute $E_0=E_k(V)+E_p(B)$.

We would like to prove that (\ref{difeqfo},\ref{icond}) is a dynamical system, that is, coordinates of points are defined for any time moment $t\in\mathbb{R}$.

Let $Z_H(t,B,V)=(X_H(t,B,V),Y_H(t,B,V))$ be a solution of (\ref{difeqfo},\ref{icond}).
The right hand side of (\ref{difeqfo}) may be regarded as a function $G_H(X,Y)=(Y,F_H(X))$,
$G_H:\mathcal{M^+}\times\mathcal{M}\longrightarrow\mathcal{M}\times\mathcal{M}$.
Each scalar component of $G_H$ is in essence a multivariate rational function, hence all of them are real analytic functions on 
$\mathcal{M^+}\times\mathcal{M}$.
So each component of $G_H$ is continuously differentiable on $\mathcal{M^+}\times\mathcal{M}$ (\cite{KP 02}, p.30). Therefore $G_H$ is locally Lipschitz on $\mathcal{M^+}$
(\cite{HS 74}, p.163).
So for any $B\in\mathcal{M^+}$, $V\in\mathcal{M}$ exists $\varepsilon>0$ such that (\cite{HS 74}, pp.162-163)
\begin{description}
\item 1) $Z_H(t,B,V)$ exists on some open interval $(t_1,t_2)$, where $t_1=t_0-\varepsilon$, $t_2=t_0+\varepsilon$;
\item 2) $Z_H(t,B,V)$ is unique.
\end{description}

The local solution defined on $(t_1,t_2)$ can be extended to a unique solution defined on maximal open interval $(\alpha,\beta)$, where
$\alpha\leq t_1$, $\beta\geq t_2$ (\cite{HS 74}, p.171). That also implies that $X_H(t,B,V)\in\mathcal{M^+}$ for all $t\in(\alpha,\beta)$.

\begin{lemma}
\label{rbound}
Suppose that $Z_H(t,B,V)$ is defined on interval $[t_0,\beta)$ which cannot be extended to the right. Suppose that $\beta<+\infty$. Then
\begin{description}
\item i) $Z_H(t,B,V)$ is bounded on interval $[t_0,\beta)$, that is, $\exists D>0\ \forall t,\ t_0\leq t<\beta$, $\|Z_H(t,B,V)\|<D$;
\item ii) $X_H(t,B,V)$ stays sufficiently far from the boundary of $\mathcal{M^+}$, that is, exists $L>0\ \forall t,\ t_0\leq t<\beta$,
$\forall(r,s)$, $r\neq s$, $\|x_r(t)-x_s(t)\|\geq L$.
\end{description}
\end{lemma}
\begin{proof}
The proof strongly relies on the constraints imposed by energy conservation law (\ref{energy}).
The proof consists of two parts. In the first part we prove proposition i) and in the second part - proposition ii).

Note that if $r\neq s$ $\|x_r(t)-x_s(t)\|>0$ for any $t\in[t_0,\beta)$.

1) Let $N(t)=\|X_H(t,B,V)\|=\sqrt{\sum\limits_{r,s=1}^m x_{rs}(t)^2}$. 
$N(t)$ is defined on $[t_0,\beta)$.
Due to Cauchy inequality (\cite{BB 71}, p.2),

$|N^\prime(t)|=\left|\displaystyle\frac{\sum\limits_{r,s=1}^m x_{rs}(t)x_{rs}^\prime(t)}{N(t)}\right|
\leq\sqrt{\sum\limits_{r,s=1}^m(x_{rs}^\prime(t))^2}=\sqrt{2E_k(t)}$.

On the other hand, due to general means inequality (\cite{BB 71}, pp.16-17),\\
$E_k(t)=E_0-E_p(t)=E_0+\frac{1}{2}\sum\limits_{\scriptstyle r,s=1\atop\scriptstyle r<s}^m(
\log\|x_r(t)-x_s(t)\|^2-h_{rs}\|x_r(t)-x_s(t)\|^2)
\leq
E_0+\sum\limits_{\scriptstyle r,s=1\atop\scriptstyle r<s}^m
\log\|x_r(t)-x_s(t)\|
=
E_0+\log\prod\limits_{\scriptstyle r,s=1\atop\scriptstyle r<s}^m
\|x_r(t)-x_s(t)\|
\leq
E_0\ +$\\
$\log\bigg(\frac{2}{m(m-1)}\sum\limits_{\scriptstyle r,s=1\atop\scriptstyle r<s}^m
\|x_r(t)-x_s(t)\|\bigg)^{\frac{m(m-1)}{2}}
=
E_0\ +$\\
$\frac{m(m-1)}{2}\log\bigg(\frac{2}{m(m-1)}\sum\limits_{\scriptstyle r,s=1\atop\scriptstyle r<s}^m
\|x_r(t)-x_s(t)\|\bigg)
\leq
E_0\ +$\\
$\frac{m(m-1)}{2}\log\bigg(\frac{2}{m(m-1)}\sum\limits_{\scriptstyle r,s=1\atop\scriptstyle r<s}^m
\Big(\|x_r(t)\|+\|x_s(t)\|\Big)\bigg)
=
E_0\ +$\\
$\frac{m(m-1)}{2}\log\bigg(\frac{2}{m}\sum\limits_{s=1}^m
\|x_s(t)\|\bigg)
\leq
E_0+\frac{m(m-1)}{2}\log\bigg(\frac{2}{\sqrt{m}}\sqrt{\sum\limits_{s=1}^m\|x_s(t)\|^2}\bigg)
=
E_0+\frac{m(m-1)}{2}\log\bigg(\frac{2}{\sqrt{m}}N(t)\bigg)
$.
Let $K=E_0+\frac{m(m-1)}{2}\log\frac{2}{\sqrt{m}}$.
So
\begin{equation}
\label{evaluation}
0\leq E_k(t)\leq K+\frac{m(m-1)}{2}\log N(t)<K+\frac{1}{2}+\frac{m(m-1)}{2}\log N(t).
\end{equation}
We have obtained a differential inequality 
\begin{equation}
\label{primedifeqen}
|N^\prime(t)|<\sqrt{2K+1+m(m-1)\log N(t)}.
\end{equation}
The  corresponding initial condition is $N(t_0)=\|B\|$, where $\|B\|=\sqrt{\sum\limits_{r,s=1}^m (B_{rs})^2}$.
Hence 
\begin{equation}
\label{difineq}
N^\prime(t)<\sqrt{2K+1+m(m-1)\log N(t)},\ N(t_0)=\|B\|.
\end{equation}
Consider initial value problem 
\begin{equation}
\label{overfunc}
N_1^\prime(t)=\sqrt{2K+1+m(m-1)\log N_1(t)},\ N_1(t_0)=\|B\|.
\end{equation}
Due to (\ref{evaluation}), $2K+1+m(m-1)\log\|B\|>0$, so (\ref{overfunc}) satisfies Lipschitz condition and has a unique solution on
$(t_0-\varepsilon,t_0+\varepsilon)$, $\varepsilon>0$. The solution of (\ref{overfunc}) is characterized by equation
\begin{equation}
t=t_0+\int\limits_{\|B\|}^{N_1(t)}\frac{dz}{\sqrt{2K+1+m(m-1)\log z}}.
\end{equation}
As $N_1\rightarrow+\infty$, the definite integral diverges and $t\rightarrow+\infty$. Therefore it is possible to extend the solution of (\ref{overfunc})
to interval $(t_0-\varepsilon,+\infty)$.
This implies that $N_1(t)$ is bounded on interval $[t_0,\beta)$.

Due to \cite{LL 69}, pp.7-8, $N_1(t)\geq N(t)$ on $[t_0,\beta)$. Note that $N(t)>0$. So $N(t)$ is bounded on $[t_0,\beta)$. 
But then $X_H(t)$ is bounded on $[t_0,\beta)$ as well.

Boundedness of $X_H(t)$ implies that exists $P>0$ such that $-E_p(t)<P$. So $0\leq E_k(t)=E_0-E_p(t)<E_0+P$. Therefore $Y_H(t)=\sqrt{2E_k(t)}$ is bounded.
Since both $X_H(t)$ and $Y_H(t)$ are bounded, $Z_H(t)$ is bounded.

2) Let $d(t)=\min\limits_{r,s}\|x_r(t)-x_s(t)\|$. Since $X_H(t)$ is bounded, exists $Z>0$ such that for all $r,s$ and $t\in[t_0,\beta)$
$\|x_r(t)-x_s(t)\|<Z$. Set $L=Z^{-\frac{m(m-1)}{2}+1}e^{-E_0}$.
Due to energy equation (\ref{energy}),\\
$-E_0\leq-E_p(t)
=
\frac{1}{2}\sum\limits_{\scriptstyle r,s=1\atop\scriptstyle r<s}^m(
\log\|x_r(t)-x_s(t)\|^2-h_{rs}\|x_r(t)-x_s(t)\|^2)
\leq$\\
$\sum\limits_{\scriptstyle r,s=1\atop\scriptstyle r<s}^m
\log\|x_r(t)-x_s(t)\|
\leq
\log d(t)+(\frac{m(m-1)}{2}-1)\log Z
$.
So $\log d(t)\geq-E_0-(\frac{m(m-1)}{2}-1)\log Z$ and $d(t)\geq L$. So $\|x_r(t)-x_s(t)\|\geq L$.
\qed
\end{proof}

\begin{lemma}
\label{lbound}
Suppose that $Z_H(t,B,V)$ is defined on interval $(\alpha,t_0]$ which cannot be extended to the left. Suppose that $\alpha>-\infty$. Then
\begin{description}
\item i) $Z_H(t,B,V)$ is bounded on interval $(\alpha,t_0]$, that is, $\exists D>0\ \forall t,\ \alpha<t\leq t_0$, $\|X_H(t,B,V)\|<D$;
\item ii) $X_H(t,B,V)$ stays sufficiently far from the boundary of $\mathcal{M^+}$, that is, exists $L>0\ \forall t,\ \alpha<t\leq t_0$,
$\forall(r,s)$, $r\neq s$, $\|x_r(t)-x_s(t)\|\geq L$.
\end{description}
\end{lemma}
\begin{proof}
Proof is essentially the same as above.
We obtain inequality (\ref{primedifeqen}).
Now $N^\prime(t)>-\sqrt{2K+1+m(m-1)\log N(t)},\ N(t_0)=\|B\|$.
Through substitution $s=-t$, $s_0=-t_0$, we define $M(s)=N(-s)=N(t)$,
where $M(s)$ is defined on interval $[s_0,-\alpha)$.
So $M^\prime(s)=-N^\prime(t)$.
We have\\ 
$-M^\prime(s)>-\sqrt{2K+1+m(m-1)\log M(s)}$, $M(s_0)=\|B\|$, yielding
$M^\prime(s)<\sqrt{2K+1+m(m-1)\log M(s)},\ M(s_0)=\|B\|$.
We have actually obtained the inequality (\ref{difineq}).
Therefore $M(s)$ is bounded on $[s_0, -\alpha)$ and $N(t)$ is bounded on $(\alpha,t_0]$.
So $X_H(t)$ is bounded on $(\alpha,t_0]$. The rest of the proof is the same as for the corresponding parts of Lemma \ref{rbound}
\qed
\end{proof}

The previous two lemmas show that no collisions are possible and no point may reach infinity in a finite time.
This is different from the classical Newtonian $n$-body problem, where {\em both} types of singularities are possible \cite{X 92}.

\begin{theorem}
\label{definterval}
The solution $Z_H(t,B,V)$ is defined on interval $(-\infty,+\infty)$.
\end{theorem}
\begin{proof}
Let us assume from the contrary that the maximal interval of the function $Z_H(t)=(X_H(t),Y_H(t))$ is $(\alpha,\beta)$, where $\beta<+\infty$. 

Since (\ref{difeqfo}) is autonomous, 
as $t\rightarrow+\beta$ either $Z_H(t)$ tends to the boundary of $\mathcal{M^+}\times\mathcal{M}$ 
or $\|Z_H(t)\|\rightarrow+\infty$, or both (\cite{HS 74}, pp.171-172).
However, due to Lemma \ref{rbound}, any of the two options is not possible, which is a contradiction. Hence $\beta=+\infty$.

For similar reason, due to Lemma \ref{lbound}, $\alpha=-\infty$.
\qed
\end{proof}

As remarked below the equation (\ref{energy}), $F_H(X)$ is real analytic on $\mathcal{M^+}$.
So a consequence of Cauchy-Kowalewskaya theorem (special case for ordinary differential equations;
\cite{KP 02}, p.42; \cite{TP 63}, pp. 555-556) is that 
for any $B\in\mathcal{M^+}$, $V\in\mathcal{M}$, the solution of (\ref{difeqfo},\ref{icond})
$Z_H(t,B,V)=(X_H(t,B,V),X_H^\prime(t,B,V))$ is real analytic at $t_0$. 
Furthermore, the consequence of Theorem \ref{definterval} is that $\forall t\in\mathbb{R}$ $X_H(t,B,V)\in\mathcal{M^+}$.
Therefore, initial conditions may be stated at any point $(t_1,Z_H(t_1,B,V))$, so Cauchy-Kowalewskaya theorem may be reapplied to state that 
$Z_H(t,B,V)$ is real analytic on $\mathbb{R}$.

So $X_H(t,B,V)$ is real analytic on $\mathbb{R}$. Hence for each $t_1\in\mathbb{R}$ $x_{ij}(t)$ may be expanded into convergent power series
$\{a_{ijn}(t_1)\}=\sum\limits_{n=0}^\infty a_{ijn}(t-t_1)^n$, where $a_{ijn}=\displaystyle\frac{x_{ij}^{(n)}(t_1)}{n!}$.
The series converge on some interval $(t_1-\varepsilon,t_1+\varepsilon)$, $\varepsilon>0$.

The power series can be compared by comparing their coefficients.
Now the possibility to compare the power series $a_{ijn}$ instead of the functions $x_{ij}$ is asserted by the following theorem.

\begin{theorem}
\label{powvfunc}
The power series $\{a_{ijn}(t_1)\}$ and $\{a_{rsn}(t_1)\}$ are equal iff the functions $x_{ij}(t)$ and $x_{rs}(t)$ are equal on $\mathbb{R}$.
\end{theorem}
\begin{proof}
If the functions $x_{ij}(t)$ and $x_{rs}(t)$ are equal, then their expansion into power series is the same.

Suppose the power series $\{a_{ijn}(t_1)\}$ and $\{a_{rsn}(t_1)\}$ are equal. Both power series are convergent on some
interval $(t_1-\varepsilon, t_1+\varepsilon)$. So the functions $x_{ij}(t)$ and $x_{rs}(t)$ are equal on $(t_1-\varepsilon, t_1+\varepsilon)$.
However as stated in \cite{KP 02}, p.14, if two functions are real analytic on an open interval $U$ and are equal on an open interval $V\subset U$ then
these functions are equal on $U$. Therefore $x_{ij}(t)$ and $x_{rs}(t)$ are equal on $\mathbb{R}$.
\qed
\end{proof}

Let us now consider some of the combinatorial properties of the differential equation (\ref{difeq1}).

Let $f_H(B,t)=X_H(t,B,V_0)$. So $f_H(B,t)$ is a real matrix whose rows are coordinates of $m$ points of the physical system at the time moment
$t\in\mathbb{R}$,
with initial configuration of points being $B$. Let $P$ be an arbitrary $m\times m$ permutation matrix. Relabeling of the coordinates of
the points does not have impact on the evolution of the system, therefore
it is quite straightforward that
\begin{equation}
\label{vienseq}
\begin{tabular}{|c|}
\hline
\\
\ $f_H(BP,t)=f_H(B,t)P$.\ \ \\
\\
\hline
\end{tabular}
\end{equation}
Quite similarly, relabeling of the points (and the vertices of the graph $H$, respectively) 
does not have effect 
on the evolution of our physical system, therefore
\begin{equation}
\label{diveq}
\begin{tabular}{|c|}
\hline
\\
\ $f_H(PB,t)=Pf_{P^THP}(B,t)$.\ \ \\
\\
\hline
\end{tabular}
\end{equation}

Equations (\ref{vienseq},\ref{diveq}) play key role in the partitioning algorithm introduced in the next section.

\section{Partitioning Algorithm}
\label{algo}
Due to Theorem \ref{reduct}, graph isomorphism problem is reducible to the automorphism partitioning of doubly
connected graphs. In this section, a partitioning algorithm A1 is described which is applied for this
type of graphs.

Let us consider the system (\ref{difeq1}), which defines the dynamics of $m$ points corresponding to a graph vertices.
It is aimed that future coordinates of the points are characterized only by dynamics of the system and are not dependent from their initial
configuration. The system is embedded into $m$-dimensional space exactly for this purpose; initial coordinates of the points are characterized 
by an $m\times m$ identity matrix $I_m$, that is, the coordinates vectors of the points initially are

\begin{tabular}{lll}
$x_1$&$=$&$(1,0,0,\dots,0,0)$\\
$x_2$&$=$&$(0,1,0,\dots,0,0)$\\
$x_3$&$=$&$(0,0,1,\dots,0,0)$\\
\multicolumn{3}{c}{$\dots\dots\dots\dots\dots\dots\dots$}\\
$x_{m-1}$&$=$&$(0,0,0,\dots,1,0)$\\
$x_m$&$=$&$(0,0,0,\dots,0,1)$.
\end{tabular}\\
The initial velocity of the points is $0$.

Hence we consider a case where the coordinates of the points initially form the identity matrix $I$ and the initial velocity of the points is $0$.
That corresponds to a system
\begin{equation}
\label{difeq2}
\left\{
\begin{array}{l}
X^{\prime\prime}=F_H(X)\\
X(0)=I\\
X^\prime(0)=O,
\end{array}
\right.
\end{equation}
where $O$ is zero matrix.

Let $A$ be a permutation matrix corresponding to an automorphism of the graph $H$. Hence $A^THA=H$.
Recall equations (\ref{vienseq}, \ref{diveq}) which express some of the properties of the function $f_H$, characterizing the future
coordinates of the system depending on the initial configuration.
By (\ref{vienseq}) and (\ref{diveq}), $f_H(I,t_0)=f_H(A^TIA,t_0)=A^Tf_{AHA^T}(I,t_0)A=A^Tf_H(I,t_0)A$.
That is,
\begin{equation}
\label{triseq}
\begin{tabular}{|c|}
\hline
\\
\ $A^Tf_H(I,t_0)A=f_H(I,t_0)$.\ \ \\
\\
\hline
\end{tabular}
\end{equation}
We say that two rows $r_1$ and $r_2$ of a matrix are permutation equivalent, if exists a permutation matrix $P$, such that $r_1P=r_2$.
Let $X_H(t)=X_H(t,I,O)=f_H(I,t)$ be the solution of (\ref{difeq2}).
A straightforward corollary of the equation (\ref{triseq}) is the following theorem.
\begin{theorem}
\label{theor1}
If two vertices of a graph $H$ are automorphism equivalent, then the corresponding two rows (and columns) in $X_H(t)$ are permutation
equivalent.
\end{theorem}
The last theorem reveals the idea of the vertices partitioning algorithm - comparison of those $m^2$ functions $x_{ij}$ which are
the solutions of the system (\ref{difeq2}).
Two vertices are in the same class of partitioning iff the corresponding rows of $X_H(t)$ are permutation equivalent.

One method to compare the functions would be numerical computation of $X_H(t_0)$ at some time moment $t_0$. Such an algorithm would be relatively fast.
However, two different functions may as well be of the same value at $t_0$. It is also not quite clear which is the precision required for such computation,
to separate two values, say, $x_{1,2}(t_0)$ and $x_{2,3}(t_0)$, which could be very close to each other and yet not equal.

However, due to Theorem \ref{powvfunc}, instead of comparing the functions $x_{ij}$ and $x_{rs}$ directly it is possible to compute $X_H(t)$ as formal power series 
and then compare coefficients of the first terms of the series.

Hence $\forall(i,j)$, $1\leq i,j\leq m$, $x_{ij}(t)=\sum\limits_{n=0}^\infty a_{ijn}t^n$, where $a_{ijn}=\displaystyle\frac{x_{rs}^{(n)}(0)}{n!}$. 
Let 
$A(n)=\left(
\begin{array}{ccc}
a_{11n} & \dots & a_{1mn}\\
\multicolumn{3}{c}{$\dots\dots\dots\dots\dots$}\\
a_{m1n} & \dots & a_{mmn}
\end{array}
\right)$,
so $X_H(t)=\sum\limits_{n=0}^\infty A(n)t^n$.

Let us solve the problem (\ref{difeq2}) in terms of formal power series.
We denote $a_{ij(2n)}$ as $a_{ij}^n$, $A(2n)$ as $A_n$, and 
seek for the solution in the form 
$x_{ij}(t)=a_{ij}=\sum\limits_{n=0}^\infty a_{ij}^n t^{2n}$, thinking of $x_{ij}$ as series with unknown coefficients.
Due to the initial conditions, $A_0=I$.

Consider the equation (\ref{difeq}).
Let us compute the $s$-th term of the left and the right side of (\ref{difeq}).
As for the $s$-th term of the series $x_{ij}^{\prime\prime}$,
\begin{equation}
\label{left}
[x_{ij}^{\prime\prime}]_s=2(s+1)(2s+1)a_{ij}^{s+1}.
\end{equation}
Recall that for two power series
\begin{equation}
\left[\bigg(\sum\limits_{n=0}^\infty a_nt^n\bigg)\bigg(\sum\limits_{n=0}^\infty b_nt^n\bigg)\right]_s=\sum\limits_{d=0}^s a_d b_{s-d},
\end{equation}
and if $a_0\neq0$, $b_s=\left[\bigg(\sum\limits_{n=0}^\infty a_nt^n\bigg)^{-1}\right]_s$ may be computed by recurrence
\begin{equation}
b_s=\displaystyle-\frac{1}{a_0}\sum\limits_{c=1}^s a_c b_{s-c},\ \mathrm{where\ }b_0=\frac{1}{a_0}.
\end{equation}
Now 
\begin{equation}
\displaystyle\big[\|x_i-x_k\|^2\big]_c=\sum\limits_{l=1}^m\sum\limits_{d=0}^c(a_{il}^d-a_{kl}^d)(a_{il}^{c-d}-a_{kl}^{c-d}).
\end{equation}
So $\big[\|x_i-x_k\|^2\big]_0=2(1-\delta_{ik})$ and 
$r_{ik}^s=\big[\|x_i-x_k\|^{-2}\big]_s$, $i\neq k$, is computed by recurrence 
\begin{equation}
r_{ik}^s=-\frac{1}{2}\sum\limits_{c=1}^s r_{ik}^{s-c}\big[\|x_i-x_k\|^2\big]_c,\ \mathrm{where\ }r_{ik}^0=\frac{1}{2}.
\end{equation}
We define $r_{ii}^s=0$.
Therefore using the right side of (\ref{difeq}), we obtain
\begin{equation}
\label{right}
[x_{ij}^{\prime\prime}]_s=\left(\sum\limits_{k=1}^m\sum\limits_{p=0}^s(a_{ij}^p-a_{kj}^p)r_{ik}^{s-p}\right)-\sum\limits_{k=1}^m h_{ik}(a_{ij}^s-a_{kj}^s).
\end{equation}
In complete analogy to $A_n$, we introduce matrices $R_n=(r_{ij}^n)$ and define ${\bf 1}$ as an $m\times m$ matrix where $\forall i,j$ ${\bf 1}_{ij}=1$.
So equations (\ref{left}) and (\ref{right}) give a doubly recurrent formula to compute the $s$-th term of the series $x_{ij}$,
\begin{equation}
\label{formula}
\begin{array}{rcl}
a_{ij}^{s+1}&=&\frac{1}{2(s+1)(2s+1)}
\Bigg(\bigg(\sum\limits_{k=1}^m\sum\limits_{p=0}^s(a_{ij}^p-a_{kj}^p)r_{ik}^{s-p}\bigg)-\sum\limits_{k=1}^m h_{ik}(a_{ij}^s-a_{kj}^s)\Bigg)\\
\vspace{0 mm}
\\
r_{ik}^s&=&-\frac{1}{2}\sum\limits_{c=1}^s r_{ik}^{s-c}\sum\limits_{l=1}^m\sum\limits_{d=0}^c(a_{il}^d-a_{kl}^d)(a_{il}^{c-d}-a_{kl}^{c-d}),
\end{array}
\end{equation}
where $A_0=I$ and $R_0=\frac{1}{2}({\bf 1}-I)$.

Our solution implies that the function's $x_{ij}(t)$ derivatives of odd order at $t=0$ vanish, $x_{ij}^{(2q+1)}(0)=0$.
The formula (\ref{formula}) is in essence an algorithm to compute $A_n$.\\
\vspace{-2 mm}\\
{\bf Algorithm A1.}
Now the vertex partitioning algorithm A1 is as follows:
\begin{description}
\item 1) compute $A_s$ for $0\leq s\leq m^2$;
\item 2) place two vertices $i$ and $j$ into one partition class if and only if $\forall s$, $0\leq s\leq m^2$, 
the $i$-th and the $j$-th row of $A_s$ are permutation equivalent.
\end{description}

The arguments why checking the first $m^2$ elements of $A_s$ should be sufficient to distinguish
between two functions that are not equal are discussed further in the section.

Now we are going to prove that A1 is a polynomial-time algorithm. For that purpose, we state the following lemma.

\begin{lemma}
\label{even}
If $a,b$ are integers such that $a\geq0$, $b>0$ then $\displaystyle\frac{(a+2b)!}{a!\,b!\,b!}$ is an even integer.
\end{lemma}
\begin{proof}
The number $\displaystyle\frac{(a+2b)!}{a!\,b!\,b!}$ is a multinomial coefficient and therefore is an integer.
If $a=0$, $\displaystyle\frac{(2b)!}{b!\,b!}=\frac{2(2b-1)!}{b!\,(b-1)!}$, so it is even.
Let $q$ be a positive integer and suppose $\displaystyle\frac{(q+2b)!}{q!\,b!\,b!}$ is even. 
Since $\displaystyle\frac{(q+1+2b)!}{(q+1)!\,b!\,b!}=\frac{(q+2b)!}{q!\,b!\,b!}+\frac{2(q+2b)!}{(q+1)!\,b!\,(b-1)!}$,
$\displaystyle\frac{(q+1+2b)!}{(q+1)!\,b!\,b!}$ is even. Hence by induction $\displaystyle\frac{(a+2b)!}{a!\,b!\,b!}$ is even.
\qed
\end{proof}

\begin{theorem}
The algorithm A1 is polynomial in time with respect to $m$.
\end{theorem}
\begin{proof}
It is straightforward to check that the number of additions and multiplications necessary to perform to compute $A_s$ is polynomial with respect to $s$.
Since $s\leq m^2$, the number of arithmetic operations is polynomial with respect to $m$.

Therefore to show that A1 is polynomial it is sufficient to prove that each arithmetical operation involving $a_{ij}^s$ and $r_{ij}^s$ can be performed
in polynomial time with respect to $s$. The numbers $a_{ij}^s$ and $r_{ij}^s$ are rational. We may assume that for each
$s$ $a_{ij}^s$ and $r_{ij}^s$ are calculated to their reduced fraction form. If a rational number is negative, we assume
that its numerator is negative. Let $a_{ij}^s=\displaystyle\frac{n_{1s}}{d_{1s}}$ and $r_{ij}^s=\displaystyle\frac{n_{2s}}{d_{2s}}$ be those reduced fractions. 
Hence we have to show that the logarithms of $|n_{1s}|,|n_{2s}|,d_{1s},d_{2s}$, i.e., the lengths of numerators and denominators, 
are bounded from above by some polynomial $P(s)$.

Let $\alpha_{ij}^s=2^s(2s)!\,a_{ij}^s$ and $\rho_{ik}^s=2^{s+1}(2s)!\,r_{ik}^s$. So $\alpha_{ij}^0$ and $\rho_{ik}^0$ are integers.
The equations (\ref{formula}) imply that
\begin{eqnarray}
\label{formula1}
&&\alpha_{ij}^{s+1}=\bigg(\sum\limits_{k=1}^m\sum\limits_{p=0}^s{2s\choose 2p}(\alpha_{ij}^p-\alpha_{kj}^p)\rho_{ik}^{s-p}\bigg)
-2\sum\limits_{k=1}^m h_{ik}(\alpha_{ij}^s-\alpha_{kj}^s),\\
\label{formula2}
&&\rho_{ik}^{s+1}=-\frac{1}{2}\sum\limits_{c=1}^{s+1}\rho_{ik}^{s-c+1}\sum\limits_{l=1}^m
\sum\limits_{d=0}^c\frac{(2(s+1))!\,(\alpha_{il}^d-\alpha_{kl}^d)(\alpha_{il}^{c-d}-\alpha_{kl}^{c-d})}{(2d)!\,(2(c-d))!\,(2(s-c+1))!}.
\end{eqnarray}
Assume that $\forall t$, $0\leq t\leq s$ , $\alpha_{ij}^t$ and $\rho_{ik}^t$ are integers. 
So by (\ref{formula1}), $\alpha_{ij}^{s+1}$ is an integer.
Consider the equation (\ref{formula2}). The numbers $\displaystyle\frac{(2s)!}{(2d)!\,(2(c-d))!\,(2(s-c))!}$ are in fact multinomial coefficients and therefore are integers.
If $c$ is an odd integer, $c=2q+1$,
\begin{eqnarray}
\lefteqn{\sum\limits_{d=0}^c\frac{(2(s+1))!\,(\alpha_{il}^d-\alpha_{kl}^d)(\alpha_{il}^{c-d}-\alpha_{kl}^{c-d})}{(2d)!\,(2(c-d))!\,(2(s-c+1))!}={}}\nonumber\\
& & {}=2\sum\limits_{d=0}^q\frac{(2(s+1))!\,(\alpha_{il}^{c-d}-\alpha_{kl}^{c-d})(\alpha_{il}^d-\alpha_{kl}^d)}{(2d)!\,(2(c-d))!\,(2(s-c+1))!}.\nonumber
\end{eqnarray}
If $c$ is even, $c=2q$,
\begin{eqnarray}
&\displaystyle\sum\limits_{d=0}^c\frac{(2(s+1))!\,(\alpha_{il}^d-\alpha_{kl}^d)(\alpha_{il}^{c-d}-\alpha_{kl}^{c-d})}{(2d)!\,(2(c-d))!\,(2(s-c+1))!}={}&\nonumber\\
&{}=\displaystyle2\sum\limits_{d=0}^{q-1}\frac{(2(s+1))!\,(\alpha_{il}^d-\alpha_{kl}^d)(\alpha_{il}^{c-d}-\alpha_{kl}^{c-d})}{(2d)!\,(2(c-d))!\,(2(s-c+1))!}&+
\displaystyle \frac{(2(s+1))!\,(\alpha_{il}^q-\alpha_{kl}^q)^2}{c!\,c!\,(2(s-c+1))!}.\nonumber
\end{eqnarray}
By Lemma \ref{even}, $\displaystyle\frac{(2(s+1))!}{c!\,c!\,(2(s-c+1))!}$ is even, therefore in both cases\\
$\displaystyle\sum\limits_{d=0}^c\frac{(2(s+1))!\,(\alpha_{il}^d-\alpha_{kl}^d)(\alpha_{il}^{c-d}-\alpha_{kl}^{c-d})}{(2d)!\,(2(c-d))!\,(2(s-c+1))!}$ is even.
So by (\ref{formula2}) $\rho_{ik}^{s+1}$ is an integer. Hence by induction $\forall s\geq0$ $\alpha_{ij}^s$ and $\rho_{ik}^s$ are integers.

However, $a_{ij}^s=\displaystyle\frac{\alpha_{ij}^s}{2^s(2s)!}$ and $r_{ik}^s=\displaystyle\frac{\rho_{ik}^s}{2^{s+1}(2s)!}$. So
$d_{1s}\leq 2^s(2s)!$\,, $d_{2s}\leq 2^{s+1}(2s)!$ and $\log d_{1s}\leq c_1(s\log s)$, $\log d_{2s}\leq c_2(s\log s)$.

On the other hand, $a_{ij}^s$ is generated by real analytic function $x_{ij}(t)$, whereas $r_{ik}^s$ is generated by a function 
$r_{ik}(t)=\displaystyle\frac{1}{\|x_i(t)-x_k(t)\|^2}$, which is real analytic for the same reason as $x_{ij}(t)$. Therefore
(\cite{KP 02}, p.15) exist constants $C>0$ and $R>0$ such that $|a_{ij}^s|\leq\displaystyle\frac{C}{R^s}$ and 
$|r_{ij}^s|\leq\displaystyle\frac{C}{R^s}$. Hence $|\alpha_{ij}^s|\leq\displaystyle\frac{C\,2^s(2s)!}{R^s}$ and 
$|\rho_{ij}^s|\leq\displaystyle\frac{C\,2^{s+1}(2s)!}{R^s}$. So $\log|n_{1s}|\leq c_3(s\log s)$ and $\log|n_{2s}|\leq c_4(s\log s)$.

Theorem is proved.
\qed
\end{proof}

By Theorem \ref{theor1}, if two vertices are automorphism equivalent, the algorithm A1 places them into one partition class.
It is however an open question what are the types of graphs the algorithm A1 produces an automorphism partition for.

The condition that a graph must be doubly connected is necessary. Suppose a graph $G$ consists of two non-isomorphic strongly regular graphs of the same 
parameters. The graph $G$ is not connected, hence it is not doubly connected. As experiments show, in this case, algorithm A1 does not give automorphism 
partition of $G$ (and $\overline{G}$ as well), which happens for essentially the same reason as why Gudkov-Nussinov algorithm fails \cite{SJC 03}.

Currently we do not have any additional satisfactory explanation, yet the following intuitive argument might be helpful.
If a graph is not connected it consists of several connected subgraphs. So intuitively, there is no enough interaction between these subgraphs
for algorithm A1 to work in this case. The situation for complement of a disconnected graph is dually the same, now {\em both} repulsion and attraction
forces are in effect between every two vertices each belonging to one of the corresponding subgraphs. Again, the interaction among the subgraphs is too 
homogeneous.

The algorithm A1 only computes the first $m^2$ coefficients of the series $a_{ij}$.
It is not formally proved whether this is sufficient. However, below a symbolic algorithm A1' devised from A1 is presented,
where first $m^2$ are proved to be sufficient to distinguish between two different functions.

Let $\alpha_{ij}^s$ be the $s$-th partial sum of $a_{ij}$, i.e., $\alpha_{ij}^s=\sum\limits_{n=0}^s a_{ij}^s t^{2s}$.
So $\alpha_{ij}^0=a_{ij}^0$.
Let $Z_s$ be the set of $s$-th partial sums of formal power series and define the convolution and deconvolution operations 
$*,\oslash:Z_s\times Z_s\longrightarrow Z_s$,
where $a_{ij}*b_{ij}$, called convolution, is the $s$-th partial sum of the multiplication
$a_{ij}b_{ij}$ and $a_{ij}\oslash b_{ij}$, called deconvolution, is the $s$-th partial sum of the polynomial division of $a_{ij}$ by $b_{ij}$.

It is straightforward to verify that (\ref{difeq}) and (\ref{formula}) imply that
\begin{equation}
\label{psum}
\alpha_{ij}^{s+1}=a_{ij}^0+\int\limits_{0}^{t}\!\!\!\int\limits_{0}^{t}\Big(\sum\limits_{\scriptstyle k=1\atop\scriptstyle k\neq i}^m
\big((\alpha_{ij}^s-\alpha_{kj}^s)\oslash\delta_{ik}^s-(\alpha_{ij}^s-\alpha_{kj}^s)h_{ik}\big)\Big),
\end{equation}
where $\delta_{ik}^s=\|\alpha_i^s-\alpha_k^s\|^2=\sum\limits_{l=1}^{m}\big((\alpha_{il}^s-\alpha_{kl}^s)*(\alpha_{il}^s-\alpha_{kl}^s)\big)$.

Now if $a_{ir}\neq a_{js}$ then $\exists k$, such that $\alpha_{ir}^k\neq\alpha_{js}^k$. To evaluate $k$, let us transform (\ref{psum})
into the following symbolical algorithm. Let $\Sigma$, $\Gamma$ be letter sequences $\{\alpha_0,\alpha_1,\dots\}$ and $\{\beta_0,\beta_1,\dots\}$,
respectively. Consider matrices $C^s$ and $D^s$ with elements in $\Sigma$ and $\Gamma$, respectively. Let 
$c_{ij}^0=\left\{
\begin{array}{ll}
\alpha_0, & i=j\\
\alpha_1, & i\neq j\\
\end{array}
\right.
$.
For any $s$, we compute $D^s$ in the following way. We define ordering on $\mathbb{N}^2$; $(k,l)<(i,j)$ iff $(k=i\ \&\ l<j)$ or $k<i$.
\begin{enumerate}
\item FORALL $i,j$, let $P_{ij}=[(c_{i1}^s,c_{k1}^s),(c_{i2}^s,c_{k2}^s),\dots,(c_{im}^s,c_{km}^s)]$ a vector of pairs;
\item FORALL $i,j$, sort $P_{ij}$ as pairs;
\item $n=0$;
\item FOR $i=1$ TO $m$\\
 \indent\quad FOR $j=1$ TO $m$\\
 \indent\quad\quad IF $\exists k,l$ such that $(k,l)<(i,j)$ and $P_{kl}=P_{ij}$ THEN\\
 \indent\quad\quad\quad $d_{ij}^s=d_{kl}^s$\\
 \indent\quad\quad ELSE\\
 \indent\quad\quad\quad $d_{ij}^s=\beta_n$\\
  \indent\quad\quad\quad $n=n+1$;
\end{enumerate}
Now $C^{s+1}$ is computed as follows.
\begin{enumerate}
\item FORALL $i,j$, let\\
\indent \quad $T_{ij}=[(c_{ij}^s,c_{1j}^s,d_{i1}^s,h_{i1}),(c_{ij}^s,c_{2j}^s,d_{i2}^s,h_{i2}),\dots,(c_{ij}^s,c_{mj}^s,d_{im}^s,h_{im})]$ a vector\\
\indent\quad of 4-tuples;
\item FORALL $i,j$, sort $T_{ij}$ as tuples;
\item $n=0$;
\item FOR $i=1$ TO $m$\\
 \indent\quad FOR $j=1$ TO $m$\\
 \indent\quad\quad IF $\exists k,l$ such that $(k,l)<(i,j)$ and $T_{kl}=T_{ij}$ THEN\\
 \indent\quad\quad\quad $c_{ij}^s=c_{kl}^s$\\
 \indent\quad\quad ELSE\\
 \indent\quad\quad\quad $c_{ij}^s=\alpha_n$\\
  \indent\quad\quad\quad $n=n+1$;
\end{enumerate}

Similarly as in Theorem \ref{theor1}, if two vertices $i$, $j$ are automorphism equivalent then the corresponding two rows
(and columns) in $C^s$ are permutation equivalent.
\\
{\bf Algorithm A1'.}
\begin{description}
\item 1) compute $C^s$ until $C^s=C^{s+1}$;
\item 2) place two vertices $i$ and $j$ into one partition class if and only if $\forall s$, $0\leq s\leq m^2$, 
the $i$-th and the $j$-th row of $C^s$ are permutation equivalent.
\end{description}

The algorithm A1' is polynomial and now checking the first $m^2$ elements is clearly sufficient:

\begin{theorem}
\label{compar}
The algorithm A1' is polynomial and the number of steps is bounded by $m^2$.
If $x_{ir}(t)$ and $x_{js}(t)$ are not equal, then $\exists k$, $0\leq k\leq m^2$, such that $c_{ir}^k\neq c_{js}^k$.
\end{theorem}
\begin{proof}
The symbolic algorithm is actually terminated as soon as for some $s$ $C^s=C^{s+1}$. Indeed, in that case for any $t>s$ $C^t=C^s$.
Further, if for some $s$ $c_{ij}^s\neq c_{kl}^s$ then $c_{ij}^{s+1}\neq c_{kl}^{s+1}$.
It can be further noted that the algorithm performs no more than $m^2$ steps, since there can be no more than 
$m^2$ mutually different elements in $C^s$ and in each step, letters from $\Sigma$ are assigned in the same order.

In a way, the symbolic algorithm mimics the arithmetical expressions of (\ref{psum}).
So if  $x_{ir}(t)\neq x_{js}(t)$ then also $\alpha_{ij}^s\neq\alpha_{kl}^s$, which in turn implies
$c_{ij}^s\neq c_{kl}^s$. But then exists $t\leq m^2$ such that $c_{ij}^t\neq c_{kl}^t$.
\qed
\end{proof}

The algorithm A1' essentially covers the algorithm A1. Indeed, if for some graph A1 gives an automorphism partition, so does A1'.
Hence the discussed dynamical algorithm is covered by a symbolical (combinatorial) algorithm, which in essence means that 
the chaotic behavior of the point-masses in the presented dynamical system does not help too much.

As experiments show, contrary to the approach presented in \cite{GN 02}, the algorithms A1 and A1' may be used to distinguish between two
strongly regular graphs with the same parameters. (By Theorem \ref{reduct}, graph isomorphism problem is reduced to the automorphism partitioning
problem of a doubly connected graph.)
However, exists a counterexample where A1' (and therefore also A1) fail.

\begin{example}
Apply the algorithm A1' to the pair of graphs A25 and B25 in \cite{M 78}, which are not isomorphic. Automorphism partition of the resulting doubly connected 
graph is not achieved and hence the solution to the graph isomorphism problem is not achieved.
\end{example}

\section{Yes-No-Don't Know Algorithm}
\label{conj}

Algorithms A1 and A1' may give two answers to the question whether two graphs are isomorphic, i.e, No or Don't Know.
If the answer is No, then with all certainty the graphs are not isomorphic.

However, it is possible to give an algorithm A2, which gives three answers: Yes (in that case the new algorithm supplies isomorphism $\gamma$ between
the two graphs), No (with all certainty the graphs are not isomorphic) or Don't Know. 

First, let us describe an algorithm which reduces the computation of $\gamma$ to the automorphism partitioning of a doubly connected graph.
A degree of a vertex $\sigma$ of a graph $G$ will be denoted as $\deg(\sigma,G)$.

{\bf Verification Algorithm.} Suppose two separate graphs $G_1=(V_1,E_1)$ and $G_2=(V_2,E_2)$ are given, $|G_1|=|G_2|=n>1$.
Without affecting generality it is possible to assume that both $G_1$ and $G_2$ are connected and have equivalent degree partitions.
Apply the reduction algorithm of Theorem \ref{reduct} to construct a graph $G_\tau=\Big(V_1\cup V_2,E_1\cup E_2\cup\big\{\{\sigma_0,\tau\}\big\}\Big)$.

Now apply automorphism partitioning algorithm on $G_\tau$, which is doubly connected.

If no $\tau$ exists, such that $\sigma_0$ and $\tau$ are in the same automorphism class of $G_\tau$, then by Lemma \ref{i_at}
the graphs are not isomorphic.

Now suppose exists $\tau$, such that $\sigma_0$ and $\tau$ are automorphism equivalent in $G_\tau$. Then by Lemma \ref{i_at} the
graphs are isomorphic.
Let $\tau_0=\tau$, $Z_0=G_\tau$, whereas the set of edges of $Z_0$ denoted as $R_0=E_1\cup E_2\cup\big\{\{\sigma_0,\tau_0\}\big\}$.
The vertex $\sigma_0$ has maximal degree in $G_1$, and so does $\tau_0$, $\deg(\sigma_0,G_1)=\deg(\tau_0,G_2)=d_0$.
Hence $\sigma_0$ and $\tau_0$ are the only vertices with degree $d_0+1$ in $Z_0$ and therefore they are the only vertices in their class
of $P(Z_0)$. 
The process which leads to the computation of explicit isomorphism from $G_1$ to $G_2$ is as follows.

Step 1. Let $\sigma_1\in V_1$ be a vertex with the largest degree in $G_1$, excluding $\sigma_0$. For every automorphism $\pi\in A(Z_0)$, such that 
$\sigma_0\pi=\tau_0$, $V_1\pi=V_2$. Hence exists $\tau_1\in V_2$ such that $\sigma_1$ and $\tau_1$ are in the same class of $P(Z_0)$.
So exists an automorphism $\pi_1\in A(Z_0)$ such that
$\forall i$, $0\leq i\leq1$, $\sigma_i\pi_1=\tau_i$ and $\tau_i\pi_1=\sigma_i$.
Construct $Z_1=(V_1\cup V_2,R_1)$ from $Z_0$ by setting $R_1=R_0\cup\big\{\{\sigma_0,\tau_1\},\{\sigma_1,\tau_0\}\big\}$. So
$\deg(\sigma_0,Z_1)=\deg(\tau_0,Z_1)>\deg(\sigma_1,Z_1)=\deg(\tau_1,Z_1)$. Since $R_0\pi_1=R_0$, $\pi_1\in A(Z_1)$.
Apply automorphism partitioning algorithm on $Z_1$.

Step 2. Now let $\sigma_2\in V_1$ be a vertex with the largest degree in $G_1$, excluding $\sigma_0,\sigma_1$. Again, since $V_1\pi_1=V_2$,
exists $\tau_2\in V_2$ such that $\sigma_2$ and $\tau_2$ are in the same class of $P(Z_1)$. Hence exists an automorphism $\pi_2\in A(Z_1)$ such that
$\forall i$, $0\leq i\leq2$, $\sigma_i\pi_2=\tau_i$ and $\tau_i\pi_2=\sigma_i$.
Construct $Z_2=(V_1\cup V_2,R_2)$ from $Z_1$ by setting $R_2=R_1\cup\big\{\{\sigma_i,\tau_{2-i}\}\ |\ 0\leq i\leq2\big\}$. So
$\deg(\sigma_0,Z_2)=\deg(\tau_0,Z_2)>\deg(\sigma_1,Z_2)=\deg(\tau_1,Z_2)>\deg(\sigma_2,Z_2)=\deg(\tau_2,Z_2)$. Since $R_1\pi_2=R_1$, $\pi_2\in A(Z_2)$.
Apply automorphism partitioning algorithm on $Z_2$.

Before the j-th step, we have obtained a graph $Z_{j-1}$ such that $\forall i$, $0\leq i\leq j-1$, the vertex $\sigma_i\in V_1$ is 1-connected to vertices 
$\{\tau_k\in V_2\ |\ 0\leq k\leq j-1-i\}$.
Given $\deg(\sigma_i,G_1)=\deg(\tau_i,G_2)=d_i$, for all $i$, $0\leq i\leq j-1$, $\deg(\sigma_i,Z_{j-1})=\deg(\tau_i,Z_{j-1})=d_i+j-i$.
Since $\forall i$, $1\leq i\leq j-1$, $d_{i-1}\geq d_i$,
$\deg(\sigma_{i-1},Z_{j-1})=\deg(\tau_{i-1},Z_{j-1})>\deg(\sigma_i,Z_{j-1})=\deg(\tau_i,Z_{j-1})$.
Furthermore, $\forall i$ $\sigma_i$ and $\tau_i$ are the only vertices with a given degree in $Z_{j-1}$, so $\{\sigma_i,\tau_i\}\in P(Z_{j-1})$
and exists automorphism $\pi_{j-1}\in A(Z_{j-1})$ such that $\forall i$, $0\leq i\leq j-1$, $\sigma_i\pi_{j-1}=\tau_i$ and $V_1\pi_{j-1}=V_2$.
Let $\sigma_j\in V_1$ be a vertex with the largest degree in $G_1$, excluding $\{\sigma_i\ |\ 0\leq i\leq j-1\}$. Since $V_1\pi_{j-1}=V_2$,
exists $\tau_j\in V_2$ such that $\sigma_j$ and $\tau_j$ are in the same class of $P(Z_{j-1})$. So exists an automorphism $\pi_j\in A(Z_{j-1})$ such that
$\forall i$, $0\leq i\leq j$, $\sigma_i\pi_j=\tau_i$ and $\tau_i\pi_j=\sigma_i$.
Construct $Z_j=(V_1\cup V_2,R_j)$ from $Z_{j-1}$ by setting $R_j=R_{j-1}\cup\big\{\{\sigma_i,\tau_{j-i}\}\ |\ 0\leq i\leq j\big\}$. So 
$\forall i$, $1\leq i\leq j$, $\deg(\sigma_{i-1},Z_j)=\deg(\tau_{i-1},Z_j)>\deg(\sigma_i,Z_j)=\deg(\tau_i,Z_j)$.
Since $R_{j-1}\pi_j=R_{j-1}$, $\pi_j\in A(Z_j)$.

The process is continued until there are no two different vertices in $V_1$ which are in the same class of $P(Z_p)$, where the number of steps $p\leq n-2$.
This bound is achieved if a complete graph $K_n$ is tested for isomorphism to a copy of itself. The computed isomorphism $\gamma$ is identified by $P(Z_p)$
because every class of $Z_p$ has exactly two vertices, one in $V_1$ and other in $V_2$.\\
\vspace{-2 mm}\\
{\bf Algorithm A2.}
For every $j$, $0\leq j\leq n-2$, $Z_j$ is doubly connected. To prove that, it is sufficient to consider $Z_{n-2}$ constructed from $G_1=G_2=K_n$.
In $\overline{Z_{n-2}}$, $\sigma_{n-1}\in V_1$ is 1-connected to every vertex in $V_2$, whereas $\tau_{n-1}\in V_2$ is 1-connected to every vertex in $V_1$.
Hence $\overline{Z_{n-2}}$ is connected. Therefore verification algorithm above can be transformed to the Yes-No-Don't Know algorithm A2 by
using partitioning algorithm A1 instead of generic automorphism partitioning algorithm. If after some step $j$ the process could not be continued for any reason 
(for example, $\sigma_j$ and $\tau_j$ are not in the same class of partition attained by applying A1 to $Z_j$), the answer is Don't Know. After the last step,
the computed bijection $\gamma$ is tested for being an isomorphism. If $\gamma$ is an isomorphism, the answer is surely Yes, otherwise it is Don't Know.

\end{document}